\pdfoutput=1
\documentclass[10pt,conference]{IEEEtran}

\usepackage{graphicx}

\usepackage{fancyhdr}
\usepackage[normalem]{ulem}
\usepackage[sort,nocompress]{cite}
\usepackage{acro}
\usepackage{booktabs}
\usepackage[enable]{easy-todo}
\usepackage{mathtools}
\usepackage{physics}
\usepackage{amsmath,amssymb,amsfonts}
\usepackage{algorithmic}
\usepackage{graphicx}
\usepackage{textcomp}
\usepackage{xcolor}
\usepackage{braket}

\usepackage[bookmarks=true,breaklinks=true,letterpaper=true,colorlinks,citecolor=blue,linkcolor=blue,urlcolor=blue]{hyperref}

\DeclareAcronym{mu}{short=MU, long=machine units}
\DeclareAcronym{dag}{short=DAG, long=directed acyclic graph}
\DeclareAcronym{artiq}{short=ARTIQ, long=advanced real-time infrastructure for quantum physics}
\DeclareAcronym{nisq}{short=NISQ, long=noisy intermediate-scale quantum}
\DeclareAcronym{dsl}{short=DSL, long=domain-specific language}
\DeclareAcronym{jit}{short=JIT, long=just-in-time}
\DeclareAcronym{rb}{short=RB, long=randomized benchmarking}
\DeclareAcronym{dds}{short=DDS, long=direct digital synthesizer}
\DeclareAcronym{dac}{short=DAC, long=digital-to-analog converter}
\DeclareAcronym{adc}{short=ADC, long=analog-to-digital converter}
\DeclareAcronym{awg}{short=AWG, long=arbitrary waveform generator}
\DeclareAcronym{afg}{short=AFG, long=arbitrary function generator}
\DeclareAcronym{fpga}{short=FPGA, long=field-programmable gate array}
\DeclareAcronym{yb171}{short=${}^{171}$Yb$^+$, long=Ytterbium 171}
\DeclareAcronym{pmt}{short=PMT, long=photomultiplier tube}
\DeclareAcronym{api}{short=API, long=application programming interface}
\DeclareAcronym{isa}{short=ISA, long=instruction set architecture}
\DeclareAcronym{rfsoc}{short=RFSoC, long=radio frequency system-on-chip}
\DeclareAcronym{rpc}{short=RPC, long=remote procedure call}
\DeclareAcronym{mw}{short=MW, long=microwave}
\DeclareAcronym{bb1}{short=BB1, long=broadband}
\DeclareAcronym{sk1}{short=SK1, long=Solovay-Kitaev}
\DeclareAcronym{spam}{short=SPAM, long=state preparation and measurement}
\DeclareAcronym{cw}{short=CW, long=continuous wave}
\DeclareAcronym{rtio}{short=RTIO, long=real-time I/O}
\DeclareAcronym{sqst}{short=SQST, long=single-qubit state tomography}
\DeclareAcronym{gst}{short=GST, long=gate set tomography}
\DeclareAcronym{1d}{short=1D, long=one-dimensional}
\DeclareAcronym{2d}{short=2D, long=two-dimensional}
\DeclareAcronym{ddb}{short=DDB, long=device database}
\DeclareAcronym{ast}{short=AST, long=abstract syntax tree}
\DeclareAcronym{rf}{short=RF, long=radio frequency}
\DeclareAcronym{mro}{short=MRO, long=method resolution order}
\DeclareAcronym{ir}{short=IR, long=intermediate representation}
\DeclareAcronym{vcd}{short=VCD, long=value change dump}
\DeclareAcronym{gpu}{short=GPU, long=graphics processing unit}
\DeclareAcronym{vliw}{short=VLIW, long=very long instruction word}
\DeclareAcronym{simd}{short=SIMD, long=single instruction multiple data}
\DeclareAcronym{hll}{short=HLL, long=high-level language}
\DeclareAcronym{ci}{short=CI, long=continuous integration}
\DeclareAcronym{hdl}{short=HDL, long=hardware description language}
\DeclareAcronym{mems}{short=MEMS, long=microelectromechanical systems}
\DeclareAcronym{ms}{short=MS, long=Molmer-Sorensen}
\DeclareAcronym{dax}{short=DAX, long=Duke ARTIQ extensions}
\DeclareAcronym{staq}{short=STAQ, long=software-tailored architecture for quantum co-design}
\DeclareAcronym{rc}{short=RC, long=red chamber}

\title{Graph-Based Pulse Representation for Diverse Quantum Control Hardware}

\author{
\IEEEauthorblockN{
Aniket~S.~Dalvi\IEEEauthorrefmark{1}\IEEEauthorrefmark{3}\IEEEauthorrefmark{4},
Leon~Riesebos\IEEEauthorrefmark{1}\IEEEauthorrefmark{3},
Jacob~Whitlow\IEEEauthorrefmark{1},
and
Kenneth~R.~Brown\IEEEauthorrefmark{1}\IEEEauthorrefmark{5}
}

\IEEEauthorblockA{
\IEEEauthorrefmark{1} Duke Quantum Center and Department of Electrical and Computer Engineering,
Duke University, Durham, NC
}
\IEEEauthorblockA{
\IEEEauthorrefmark{5} Department of Physics and Department of Chemistry,
Duke University, Durham, NC
}
\IEEEauthorblockA{
\IEEEauthorrefmark{3}Authors~contributed~equally
}
\IEEEauthorblockA{
\IEEEauthorrefmark{4}Email: aniketsudeep.dalvi@duke.edu
}
}

\newcommand\blfootnote[1]{%
  \begingroup
  \renewcommand\thefootnote{}\footnote{#1}%
  \addtocounter{footnote}{-1}%
  \endgroup
}

\begin{document}
\date{}
\maketitle

\thispagestyle{empty}

\begin{abstract}
Pulse-level control of quantum systems is critical for enabling gate implementations, calibration procedures, and Hamiltonian evolution which fundamentally are not supported by the traditional circuit model. This level of control necessitates both efficient generation and representation. In this work, we propose \texttt{pulselib}~---~a graph-based pulse-level representation. A graph structure, with nodes consisting of parametrized fundamental waveforms, stores all the high-level pulse information while staying flexible for translation into hardware-specific inputs. We motivate \texttt{pulselib} by comparing its feature set and information flow through the pulse-layer of the software stack with currently available pulse representations. We describe the architecture of this proposed representation that mimics the abstract syntax tree (AST) model from classical compilation pipelines. Finally, we outline applications like trapped-ion-specific gate and shelving pulse schemes whose constraints and implementation can be written and represented due to \texttt{pulselib}'s graph-based architecture.
\end{abstract}

% Content
\acresetall
\section{Introduction}

\blfootnote{Parts of this paper, particularly Sections~\ref{sec:pulse:arch}, \ref{sec:pulse:transform}, and \ref{sec:pulse:implementation}, have been taken from  Leon Riesebos' PhD dissertation~\cite{riesebosphd}}
Pulse-level access to quantum systems through a convenient interface is becoming an increasingly important feature because 
programming these systems at the pulse level enables users to experiment with new pulse shapes, express low-level calibration experiments, and optimize quantum programs beyond the discrete gate model, as described in~\cite{shi2019optimized, gokhale2020optimized, werschnik2007quantum}. It also enables pulse-level access for analog quantum computing applications like simulations~\cite{Whitlow2023, Feng2023, kang2024trappedion}. 
A key feature of pulse-level access is the thin abstraction layer between the user and pulse-generating equipment such that programmers do not need to know the details of the control hardware.
Various libraries for pulse-level programming of quantum systems already exist, including Qiskit pulse~\cite{alexander2020qiskit, mckay2018qiskit}, Q-CTRL~\cite{boulder_opal1}, Pulser~\cite{silverio2022pulseropensource}, and JaqalPaw~\cite{lobser2021jaqalpaw}.
% which are all embedded into the Python language.
Unfortunately, existing pulse-level programming methods are often quantum-technology specific, pulse generation hardware specific, or semantically limited to express pulses across applications. 
In this work, we develop an architecture for pulse descriptions that is technology-agnostic, target-independent, and easy to parameterize or transform.
% We propose this pulse description to be used as an \ac{ir} to unify the pipeline from high-level pulse representation to pulse generation hardware platforms.

% We focus on pulse descriptions for unitary quantum operations with an abstract device model to express sequential and parallel pulse outputs.
Sample-based representations can represent arbitrary pulses but have a low information density and are not target-independent. Pulses that are transformed into samples lose high-level information about the original waveform, such as the frequencies in the signal.
Conversely, directly storing high-level pulse data, such as frequencies and phases, often leads to inflexible solutions that do not adapt to complex pulse shapes and modulation techniques such as described in~\cite{leung2018robust, kang2021batch, kang2022designing}.
Techniques for pulse parameterization and insertion of calibration parameters are crucial, but current implementations do not allow analysis or transformations of pulse descriptions before all parameters are concrete.
% We aim to integrate pulse-level control into a programming environment that includes universal classical and discrete quantum operations.

In this work, we present a graph-based architecture for pulse descriptions. Our underlying structure enables compact pulse descriptions and retains high-level information on complex pulses and modulation schemes by including arithmetic operations in the graph. Further, phase synchronization information for pulses is explicitly stored in the graph too. Pulse parameterization and insertion of calibration parameters are achieved using variable nodes that can be substituted after graph construction.
We introduce infrastructure for pulse schedules to allow pulses across channels to be scheduled relative to each other using notions of parallel and sequential ordering.
In our environment, recursive graph algorithms make pulse analysis and transformations possible, even before variables are substituted. Graph descriptions of pulses can eventually be rendered to samples. Alternatively, pulse properties can be extracted from the graph on a higher abstraction level using graph algorithms such as maximal munch~\cite{cattell1980automatic}. This allows the representation to be transformed to target any application or pulse generation hardware.

The major contributions of this work are as follows:
\begin{enumerate}
    \item We motivate \texttt{pulselib} by introducing the concepts of pulse \emph{creation, representation,} and \emph{realization} as phases within the pulse-level software stack.
    \item We highlight the graph-based architecture of \texttt{pulselib}. It uses a \ac{dag} design built out of fundamental scalars and waveforms.
    \item We describe the accompanying architecture around the graphs~---~schedules, transformations, and pipelines. Schedules allow for pulses across channels to be appropriately scheduled relative to each other. Transformers can be used to visit nodes and transform the graph to a desired format. Finally, pipelines allow for multiple transformers to alter the graph sequentially.
    \item We demonstrate the utility of \texttt{pulselib} by using it to represent trapped-ion pulse schemes involving complex phase synchronization schemes.
\end{enumerate}

\section{Motivation}
\label{sec:motivation}

The current landscape in quantum computing contains a handful of pulse-level representations \cite{pulser, alexander2020qiskit, landahl2020jaqal, quantummachinesQUACode}. We motivate the addition of another one, i.e., \texttt{pulselib}, by analyzing the information flow in these representations and outlining the required feature set from pulse-level representations.

The pulse layer of a quantum computing stack can be divided into three phases~---~creation, representation, and realization. The \emph{creation} phase refers to the API, syntax, and semantics of how the pulse is created. The creation could happen explicitly by the user, as in the case of optimal pulse control experiments, calibration, and analog quantum computing, or could be implicit, for example when gates are implicitly converted to pulses before being executed on the system. \emph{Representation} refers to how the pulse is represented in memory after creation. The representation of pulses in memory can be broadly categorized into parametric representations and sample-based representations. The latter stores a pulse as a list of samples at a given sample rate, while the former stores the pulse definition as a data structure with parameters like its duration, phase, frequency, and amplitude. Finally, \emph{realization} refers to converting the pulse representation to one that is used to by the underlying hardware to synthesize the pulse. This could be a series of values sent to an \ac{awg} which generates a pulse by sampling them at a given rate, or some parameters assigned to a register that a \ac{dds} uses to generate a waveform. Most pulse representations encapsulate the \emph{creation} and the \emph{representation} layers.

Given these phases, a pulse representation can be evaluated by its ability to provide semantic ease to create pulses, define arbitrary pulse schemes, and retain maximum information about the pulse until realization. During the creation phase most amount of high-level information is available, and the least during the realization phase. Considering this loss of information through the phases, it is desirable for the \textit{representation} phase to retain information before being lowered to the realization phase. Allowing creation at an earlier stage where not all information is available means we can describe abstract pulses. As a result, more information needs to be retained in the representation. Furthermore, having more information in the representation allows for a new technique: transformation of the pulse during the representation phase.

A sample-based representation allows for maximum flexibility to represent an arbitrary pulse. However, the size of the list of samples scales linearly with the duration of the pulse, making it memory inefficient. Also, the sample rate is chosen at creation thereby immediately making it hardware-specific. The semantics for the creation of a sample-based representation can be parametric which holds high-level information about the pulse, however, when the pulse is lowered to a list of samples this high-level information is difficult to extract and the context of that information is lost. A parameter-based representation, however, allows for user-friendly semantics for pulse creation and also retains information in the representation phase. This is because the pulse information stored in memory is still parametric and not reduced to a lower representation like a list of samples. A parametric pulse representation, however, limits the ability to describe completely arbitrary pulses. This is where we motivate \texttt{pulselib}, a graph-based parametric pulse representation. It efficiently stores pulse information at scale, retains high-level information until the pulse needs to be realized, and pushes the limit towards arbitrary pulse representation as required by current and future quantum systems. We elucidate this by comparing \texttt{pulselib}'s feature set to those of other parametric pulse representations shown in Table~\ref{tab:features}.

\begin{table}
    \centering
    \caption{Table comparing features between relevant parametric pulse representations. The + or - indicates a subtlety in the Yes (Y) or No (N) designation and is expanded on in the text.}
    \begin{tabular}{@{}llll@{}}
        \toprule
        Feature & Pulser & Qiskit Pulse & \texttt{pulselib}\\
        \midrule
        Publicly Available          & Y    & Y   & Y   \\
        Parametric Representation   & Y    & Y   & Y   \\
        Variables                   & Y    & Y   & Y   \\
        Device Agnostic             & N    & Y   & Y   \\
        Arbitrary                   & N    & Y   & Y   \\
        Scheduling                  & Y-   & Y   & Y   \\
        Dynamic Schedules           & N    & Y   & Y+  \\
        Interpolated Waveform       & Y    & N   & Y   \\
        Graph Based                 & N    & N+  & Y   \\
        Phase Synchronization       & N    & N   & Y   \\
        \bottomrule
    \end{tabular}
    % \textit{Y-} for Pulser's scheduling feature indicates that although it has basic scheduling capabilities because of \texttt{delay} statement, it lacks the semantics for relative pulse scheduling. \textit{Y+} for \texttt{pulselib}'s dynamic schedule means that not only can the pulse in a schedule be changed, but the contents of the pulse can also be altered. \textit{N+} for Qiskit Pulse's graph-based structure conveys that the representation is not inherently graph based, but has a function that allows the user to up convert it to a graph. }
    \label{tab:features}
\end{table}

Table~\ref{tab:features} compares \texttt{pulselib} to other parametric pulse representations~---~Pulser~\cite{pulser} and Qiskit Pulse~\cite{alexander2020qiskit}. Pulser is a representation that was developed for writing and simulating pulses for neutral atom quantum systems. This work, however, focuses on universal, device-agnostic pulse representations that may be used with any underlying hardware. Qiskit Pulse, which is device agnostic, originally started as a representation that allowed users to create pulses using parameters but internally represented them as samples. However, with the addition of the Qiskit Symbolic Pulse~\cite{ibmSymbolicPulseQuantum}, it can now be classified as a parametric pulse representation. Pulser contains a limited number of pre-defined base waveforms and does not allow for arbitrary pulses to be created, further limiting its general use case. Qiskit Pulse allows users to use SymPy~\cite{sympy} to write functions that could define arbitrary pulses. \texttt{Pulselib}'s graph-based representation contains some commonly used waveforms and allows users to extend these to make custom waveforms. These waveforms can be combined to make arbitrary waveforms for quantum systems. All three pulse representations being compared allow for pulse parameters to be variables. This allows for certain parameters to be substituted just before the pulse is realized and converted to a hardware-specific representation. This is another advantage of parametric representations over a list of samples. The graph structure of \texttt{pulselib} allows for a simple graph traversal to substitute these variables. Although Qiskit Pulse allows their representation to be up-converted to a graph, the lack of an inherent graph structure limits its usefulness. This is why Table~\ref{tab:features} marks this feature as \textit{N+}. The lack of an architectured inheritance structure for the nodes in this up-converted graph limits its utility for optimization, transformation, and extension. These features are available in \texttt{pulselib} because the inherent \ac{ast}-like graph architecture of \texttt{pulselib} results in each pulse being composed of a set of base nodes. Lastly, semantically, \texttt{pulselib} can be used as a pulse-level \ac{dsl}, but its \ac{ast}-like structure also allows it to be used as an intermediate representation in the software stack.

Individual pulses are insufficient to control quantum systems, they need to be part of a larger pulse schedule. Pulse representations, therefore, need to support schedule descriptions. Pulser allows users to add pulses with delays between them. This creates simple schedules but is semantically limited for complex pulse schemes. Consequently, Table~\ref{tab:features} denotes this as \textit{Y-}. Qiskit pulse and \texttt{pulselib} allow for more advanced scheduling with the ability to have pulses scheduled relative to each other. They also allow for schedules to be dynamic, i.e., pulses may be added or substituted into a schedule anytime post-creation and before realization. \texttt{pulselib} however, takes this further by allowing the contents of the pulses in the schedule to be transformed, thereby marking it as \textit{Y+} in Table~\ref{tab:features}. This is again a product of the graph architecture. Finally, phase synchronization is a vital aspect of pulse programming. Pulses need to be applied at the right phase to accurately alter the state of qubits. Although Pulser and Qiskit Pulse allow users to specify the phase of a waveform and perform phase-shift operations, information about phase synchronization between pulses is not explicitly captured or represented. Complicated phase synchronization in these representations cannot be expressed and has to be incorporated using phase-shift operations through user-defined, sometimes extensive, calculations during the creation phase. \texttt{pulselib}'s architecture provides specialized \textit{clock} waveforms to allow explicit representation of phase synchronization between pulses. This makes it the only pulse representation to provide explicit support for phase synchronization in the representation, with all the phase calculations handled by the representation. Section~\ref{sec:pulse:applications} describes real examples of such pulses and their corresponding \texttt{pulselib} representation.
The following section describes \texttt{pulselib}'s architecture, and how it enables the above-mentioned features.
% \acresetall
% \input{content/30_background}
\section{Architecture}
\label{sec:pulse:arch}

\begin{figure}
    \centering
    \includegraphics[width=\linewidth]{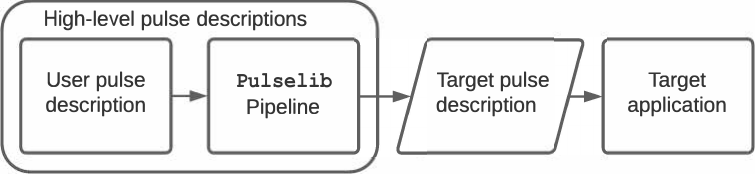}
    \caption{The envisioned workflow from graph-based user pulse descriptions to a target application.}
    \label{fig:pulse:overview}
\end{figure}

The target-independent architecture for our pulse representation is based on a set of \emph{waveforms} with one or more \emph{parameters}. Waveforms and parameters can be represented by \emph{nodes} and relations between them as labeled directed edges, which results in a \ac{dag}.
Our architecture focuses on the representation of finite duration pulses where its duration and parameters are known when the pulse is realized, but not necessarily at the time when the pulse is created.
Once we express pulses using \acp{dag}, we use graph algorithms to validate and transform the pulses and apply techniques derived from compilers and \acp{ir}. We can easily transform pulses to a format that can be realized on a target hardware for the target application. This architecture serves to create arbitrary waveforms, carry maximum high-level information from the creation to the representation, while still allowing flexible realization to the underlying hardware and application. Figure~\ref{fig:pulse:overview} shows a schematic overview of the envisioned workflow.

% \todo{note that pulseIR has seen continued development and new scalar/waveform types (and other features) should be added. one example is the polynomial or the special class hierarchy for Zero and Const waveforms.}

% \subsection{Node Types}
\subsection{Scalars}

Scalars are nodes that represent a single value without a time component. The most basic scalar is a \emph{number}~(Num), which represents an integer or float value. A number has no unit by itself, and the unit it represents depends on the relation to its super-node (i.e. parent node). A number node can have multiple super-nodes and is by definition a leaf node.

\emph{Variable} scalars~(Var) represent numbers of which the value will be \emph{substituted} later. A variable node can be used for pulse parameterization or the insertion of calibration parameters. Each variable scalar has a key that is used for substitution. To substitute a variable, we provide a key-value mapping to the node, and the value of the node is now substituted using the key and the mapping.

\begin{figure}
    \centering
    \includegraphics[width=1\linewidth]{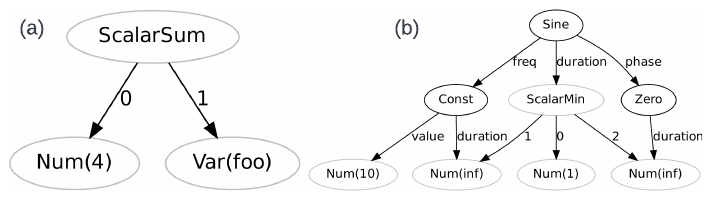}
    \caption{(a) The graph for the sum of a number and a variable with key ``foo.'' Numbers at the edges represent the operand order. (b) The graph of a sine waveform with a static frequency and phase.}
    \label{fig:pulse:scalar_sum}
\end{figure}

We might want to perform basic math operations on scalars, but the presence of variables does not allow us to evaluate the value of such operations until all variables involved are substituted. To allow math operations on scalars without evaluating the actual values, we introduce \emph{operator nodes}. A scalar operator node represents a math operation on a set of scalar nodes referred to as \emph{items}. In the \ac{dag} representation, items are considered sub-nodes of the operator node. Scalar operators include the sum, product, subtract, divide, unary minus, min, and max.
An example \ac{dag} with a sum of a number and a variable is shown in Figure~\ref{fig:pulse:scalar_sum}~(a). The labels at the edges indicate the order of the operands.
With scalar operator nodes, we can express basic math on numbers and variables using a \ac{dag} without evaluating the actual values. These variable and scalar operator nodes allow users to logically describe pulses with unknown parameters or those with an arithmetic combination of multiple parameters in the creation phase, while continuing to retain this information in the representation phase.

% \begin{table}
% \centering
% \caption{List of scalar operators and their properties.}
% \begin{tabular}{@{}llr@{}}
% \toprule
% Scalar operator & Operator    & Number of items \\
% \midrule
% ScalarSum       & Sum         & 1..n            \\
% ScalarProduct   & Product     & 1..n            \\
% ScalarSub       & Subtract    & 1..n            \\
% ScalarDiv       & Divide      & 1..n            \\
% ScalarNeg       & Unary minus & 1               \\
% ScalarMin       & Minimum     & 1..n            \\
% ScalarMax       & Maximum     & 1..n            \\
% \bottomrule
% \end{tabular}
% \label{tab:pulse:scalar_operators}
% \end{table}

\subsection{Waveforms}

Waveforms are nodes that represent a time-dependent value; they always have a duration. Each waveform is defined in the domain $[t_\textit{start}, t_\textit{start} + d)$ where $t_\textit{start}$ represents the start time of the waveform and $d$ is the scalar duration of the waveform. Outside of its domain, a waveform has a value of $0.0$.
Waveforms are allowed to have infinite duration.
Different waveform types are represented by different nodes where each has its parameters in addition to a duration parameter. Such parameters can be scalars or \textit{other waveforms} to allow parameter modulation. In a \ac{dag}, parameter relations are represented by directed and labeled edges between the waveform and the parameter node.
The waveforms currently supported by \texttt{pulselib} are~---~constant, zero (constant pulse of zero amplitude), ramp, triangle, Gaussian, clock (node representing reference clock for phase synchronization), sine, frequency-modulate sine, phase-modulated sine, polynomial (time domain polynomial waveform), and power (time domain power function).
A waveform with waveform parameters $p_0, \dots, p_{n-1}$ will have a duration of $\min(d_\textit{waveform}, d_{p_0}, \dots, d_{p_{n-1}})$ where $d_\textit{waveform}$ is the configured duration of the waveform and $d_{p_0}, \dots, d_{p_{n-1}}$ are the durations of the waveform parameters. Hence, a waveform is only defined within a domain where all its parameters are defined. 
Arbitrary waveforms can be represented by creating new waveform types, but such waveforms, if not designed thoughtfully using the nodes and parameters structure, will not leverage the graph structure well.

% \todo{I'm undecided on this table. In one sense its clear, but its sort of not important. Can we just say we support all the common pulse shapes e.g. Gauss, Sin, Ramps, Const etc.}

% \begin{table}
% \centering
% \caption{A list of waveforms and parameters (excluding the duration parameter).}
% \begin{tabular}{@{}lll@{}}
% \toprule
% Waveform & Parameters          & Type     \\
% \midrule
% Zero     &                     &          \\
% Const    & value               & Scalar   \\
% Ramp     & slope               & Waveform \\
%          & intercept           & Waveform \\
% Triangle & peak                & Waveform \\
%          & center              & Waveform \\
%          & slope               & Waveform \\
% Gauss    & peak                & Waveform \\
%          & mu                  & Waveform \\
%          & sigma               & Waveform \\
% Clock    & freq                & Scalar   \\
%          & phase               & Scalar   \\
% Sine     & freq                & Waveform \\
%          & phase               & Waveform \\
%          & ref\_clk (optional) & Clock    \\
% SineFM   & carrier             & Clock    \\
%          & modulation          & Waveform \\
%          & phase               & Waveform \\
% SinePM   & carrier             & Clock    \\
%          & modulation          & Waveform \\
% \bottomrule
% \end{tabular}
% \label{tab:pulse:waveforms}
% \end{table}

% The simplest waveform type is zero, which represents a constant value of zero for a given duration. Besides the duration, the zero waveform has no parameters.

For example, constant waveforms, which represent a constant value within its domain, are configured by a scalar value parameter. 
% The \ac{dag} of a constant waveform is shown in Figure~\ref{fig:pulse:wave_const}. 
Other waveforms, such as the sine waveform, have waveform parameters that can change their value over time.
Static parameters are now represented by constant waveforms. Figure~\ref{fig:pulse:scalar_sum}~(b) shows the \ac{dag} of a sine waveform with a fixed frequency and phase over time. Note that the \ac{dag} shows that sine duration is defined by the minimum of its parameters' durations and the configured duration of the sine waveform using a scalar operator.

% \begin{figure}
%     \centering
%     \includegraphics[width=0.6\linewidth]{fig/wave_const.gv-crop.pdf}
%     \caption{The graph of a constant waveform with a scalar value and a scalar duration.}
%     \label{fig:pulse:wave_const}
% \end{figure}

% \begin{figure}
%     \centering
%     \includegraphics[width=\linewidth]{fig/wave_sine.gv-crop.pdf}
%     \caption{The graph of a sine waveform with a static frequency and phase.}
%     \label{fig:pulse:wave_sine}
% \end{figure}

\begin{figure}
    \centering
    \includegraphics[width=\linewidth]{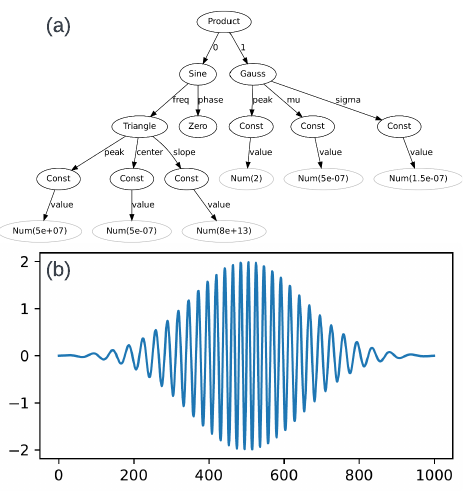}
    \caption{Traingle frequency-modulated waveform with added Gaussian amplitude modulation sampled at 1~GHz.}
    \label{fig:pulse:wave_sine_am}
\end{figure}

A powerful feature of our architecture is that waveform parameters can be modulated by using other waveforms as parameters. For example, the \ac{dag} for a sine waveform with a triangle-modulated frequency is shown as operand 0 of the product operator in Figure~\ref{fig:pulse:wave_sine_am}~(a), with duration nodes omitted for clarity. 
% Its corresponding triangle component is found in Figure \ref{fig:pulse:wave_triangle_plot} and the aggregate modulated pulse is found in Figure \ref{fig:pulse:wave_sine_fm_plot}. 
The semantics of \texttt{pulselib} seamlessly allow for this nested pulse creation, while the graph architecture holds this information in its representation in memory and can easily render it as an analytic pulse or sample from it during realization.
% The duration nodes were omitted from the graph to increase its readability, and we will do the same for most graphs onwards.
% The frequency of the sine waveform starts at 10~kHz and ramps up linearly to 50~kHz before ramping back to 10~kHz. We have plotted the values of the triangle and the sine waveforms in Figure~\ref{fig:pulse:wave_triangle_plot} and~\ref{fig:pulse:wave_sine_fm_plot}, respectively. Both plots have a duration of one microsecond using a sample rate of 1~GHz.
% Note that the frequency-modulated sine wave in Figure~\ref{fig:pulse:wave_sine_fm_plot} is computed using a phase accumulator to prevent discontinuities.

% \begin{figure}
%     \centering
%     \includegraphics[width=\linewidth]{fig/wave_sine_fm.gv-crop.pdf}
%     \caption{The graph of a sine waveform with a triangle modulated frequency from 10 to 50~kHz.}
%     \label{fig:pulse:wave_sine_fm}
% \end{figure}

% \begin{figure}
%     \centering
%     \includegraphics[width=\linewidth]{fig/wave_triangle.pdf}
%     \caption{The rendered triangle waveform from Figure~\ref{fig:pulse:wave_sine_fm}.}
%     \label{fig:pulse:wave_triangle_plot}
% \end{figure}

% \begin{figure}
%     \centering
%     \includegraphics[width=\linewidth]{fig/wave_sine_fm.pdf}
%     \caption{The rendered sine waveform with a triangle modulated frequency from Figure~\ref{fig:pulse:wave_sine_fm}.}
%     \label{fig:pulse:wave_sine_fm_plot}
% \end{figure}

Similar to scalars, we would like to perform time-wise operators on waveforms. Hence, we introduce waveform operators similar to the operator nodes for scalars but instead applied to waveforms. Waveform operators include sum, product, subtract, divide, and unary minus. 
% The unary minus waveform operator only supports one item while other waveform operators support one or more items, similar to scalar operators.
The duration of waveform operators depends on the items contained in the operator. 
% We would like to recall that the value of a waveform outside its domain is always $0.0$. A product or division containing zero is nonsensical. 
The duration of products or divisions of waveforms is the minimum duration of their items.
For the sum and subtract operators, the duration of the waveform operator is the maximum duration of its items. We agree that for the sum and subtract operators, both the minimum and maximum duration of its items are valid choices. We chose the maximum as we think this will increase the usability for these operators in practice.

We can use the product operator to create a sine wave with amplitude modulation. Figure~\ref{fig:pulse:wave_sine_am}~(a) is a triangle frequency-modulated sine wave but with added Gaussian amplitude modulation using the product operator. The resulting waveform is plotted in Figure~\ref{fig:pulse:wave_sine_am}~(b) with a sample rate of 1~GHz.
% used in Figure~\ref{fig:pulse:wave_sine_fm}.

% \begin{figure}
%     \centering
%     \includegraphics[width=\linewidth]{fig/wave_sine_am.pdf}
%     \caption{The rendered sine waveform with frequency and amplitude modulation from Figure~\ref{fig:pulse:wave_sine_am}.}
%     \label{fig:pulse:wave_sine_am_plot}
% \end{figure}

Sequence operators are used to concatenate waveforms. The sequence operator has zero or more items, and its duration is the sum of the duration of all its items.
The sequence operator can be used to sequentially order waveforms and has applications ranging from building longer waveforms to creating complex modulation patterns. An important side-effect of the sequence operator is the shifted starting time of waveforms. Waveforms with a phase could be affected by a shifted starting time while other waveforms are insensitive to shifts in starting time.
% A list of waveform operators and their properties is shown in Table~\ref{tab:pulse:waveform_operators}.

% \begin{table}
% \centering
% \caption{List of waveform operators and their properties. $d_x$ represents the duration of item $x$.}
% \small
% \begin{tabular}{p{1.5cm} p{2cm} p{1cm} p{2cm}}
% \toprule
% Waveform operator & Operator    & Number of items & Duration                           \\
% \midrule
% Sum               & Sum         & 1..n            & $\max(d_0, \dots, d_\textit{n-1})$ \\
% Product           & Product     & 1..n            & $\min(d_0, \dots, d_\textit{n-1})$ \\
% Sub               & Subtract    & 1..n            & $\max(d_0, \dots, d_\textit{n-1})$ \\
% Div               & Divide      & 1..n            & $\min(d_0, \dots, d_\textit{n-1})$ \\
% Neg               & Unary minus & 1               & $d_0$                              \\
% Sequence          & Concatenate & 0..n            & $\sum_{x=0}^{n-1} d_x$             \\
% \bottomrule
% \end{tabular}
% \label{tab:pulse:waveform_operators}
% \end{table}

\begin{figure}
    \centering
    \includegraphics[width=\linewidth]{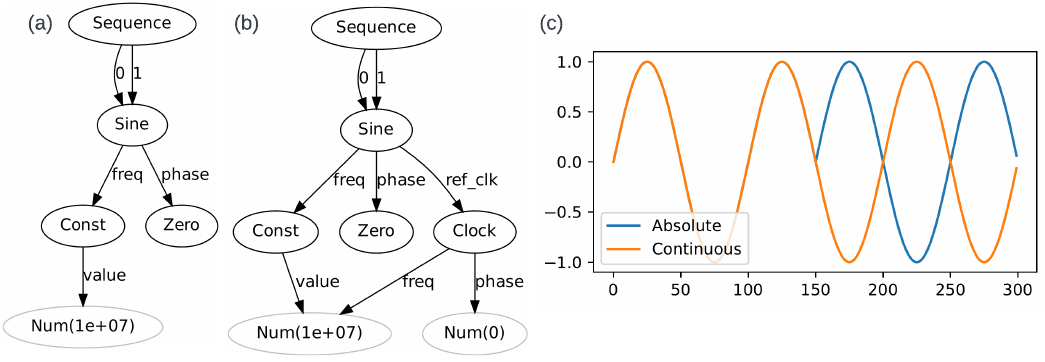}
    \caption{The graphs and renders of two sequential waveforms with absolute and continuous phase mode.}
    \label{fig:pulse:wave_sine_phase_mode_plot}
\end{figure}

We introduce two phase modes for waveforms with a phase: absolute and continuous. Phase mode absolute indicates that the phase of a waveform, in radians, is insensitive to the waveform's start time and only depends on the phase parameter of the waveform. Alternatively, phase mode continuous indicates that the phase of a waveform is time-dependent. The initial phase offset is derived from a reference clock, and the phase parameter of the waveform is relative to the initial phase offset.
The reference clock is a clock waveform with a scalar frequency and phase from which we can easily calculate the phase offset at any start time.
The usage of a reference clock allows our architecture to support a wide range of phase synchronization methods, including situations where phase tracking is performed at a different frequency than the waveform itself. \texttt{Pulselib} also includes a clock sequence~---~a sequence of reference clock waveforms each with their own durations. The clock sequence node allows for time-dependent phase accumulation, where the phase between waveforms is synchronized to a clock that accumulates phase at a time-dependent rate. These clock and clock sequence nodes allow users to capture phase synchronization while creating pulse sequences, and their representation and implementation in this graph structure facilitates implicit calculation and the transfer of this information and  when realizing the pulse on hardware.
Figure~\ref{fig:pulse:wave_sine_phase_mode_plot}~(a)-(b) shows the graphs of two identical sine waves with absolute and continuous phase modes. A plot of both waveforms is shown in Figure~\ref{fig:pulse:wave_sine_phase_mode_plot}~(c), showing a duration of 0.3~microseconds using a sample rate of 1~GHz.
In the left half of the plot, both waveforms overlap. In the right half of the plot, we see a clear difference between the absolute and continuous phase modes. Some real applications of trapped-ion pulse sequences requiring the clock and the clock sequence nodes are discussed in Section~\ref{sec:pulse:applications}.

% \begin{figure}
%     \centering
%     \subfloat[Absolute phase mode\label{fig:pulse:wave_sine_abs}]{
%         \centering
%         \includegraphics[scale=0.45]{fig/wave_sine_abs.gv-crop.pdf}
%     }
%     \hfill
%     \subfloat[Continuous phase mode\label{fig:pulse:wave_sine_const}]{
%         \centering
%         \includegraphics[scale=0.45]{fig/wave_sine_cont.gv-crop.pdf}
%     }
%     \caption{The graphs of two sequential waveforms with absolute and continuous phase mode.}
%     \label{fig:pulse:wave_sine_phase_mode}
% \end{figure}

\begin{figure}
    \centering
    \includegraphics[width=\linewidth]{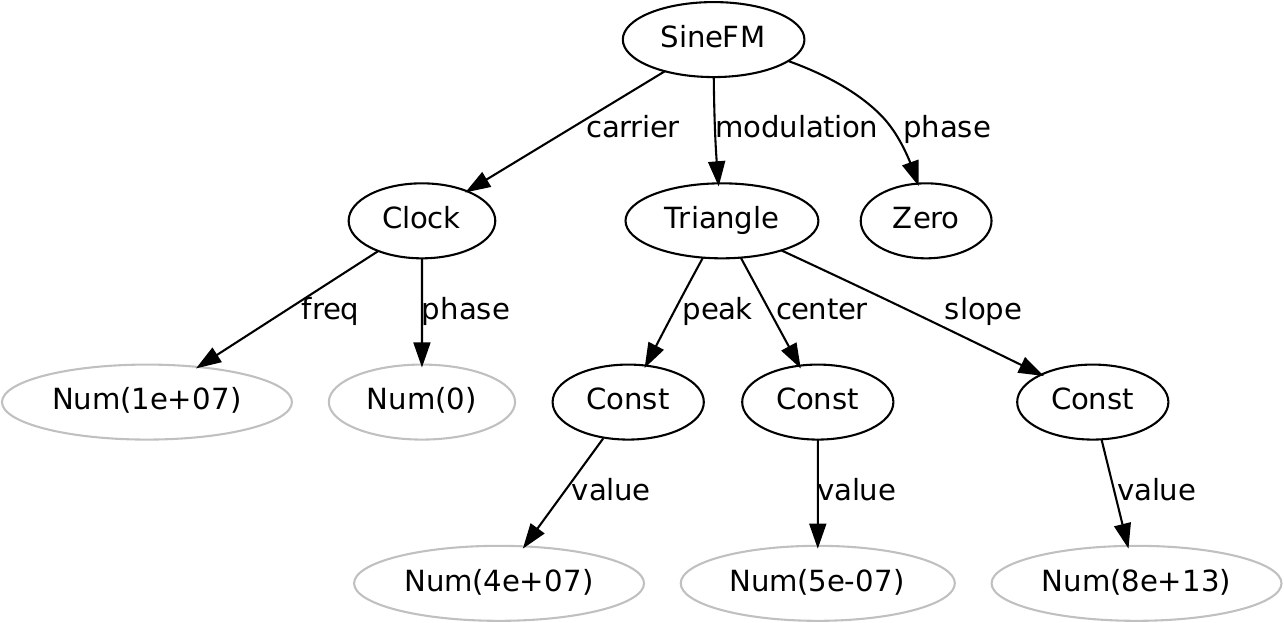}
    \caption{The graph of a frequency-modulated sine waveform
    % , similar to Figure~\ref{fig:pulse:wave_sine_fm}, 
    based on a 10~MHz carrier frequency.}
    \label{fig:pulse:wave_sinefm}
\end{figure}

\begin{figure}
    \centering
    \includegraphics[width=\linewidth]{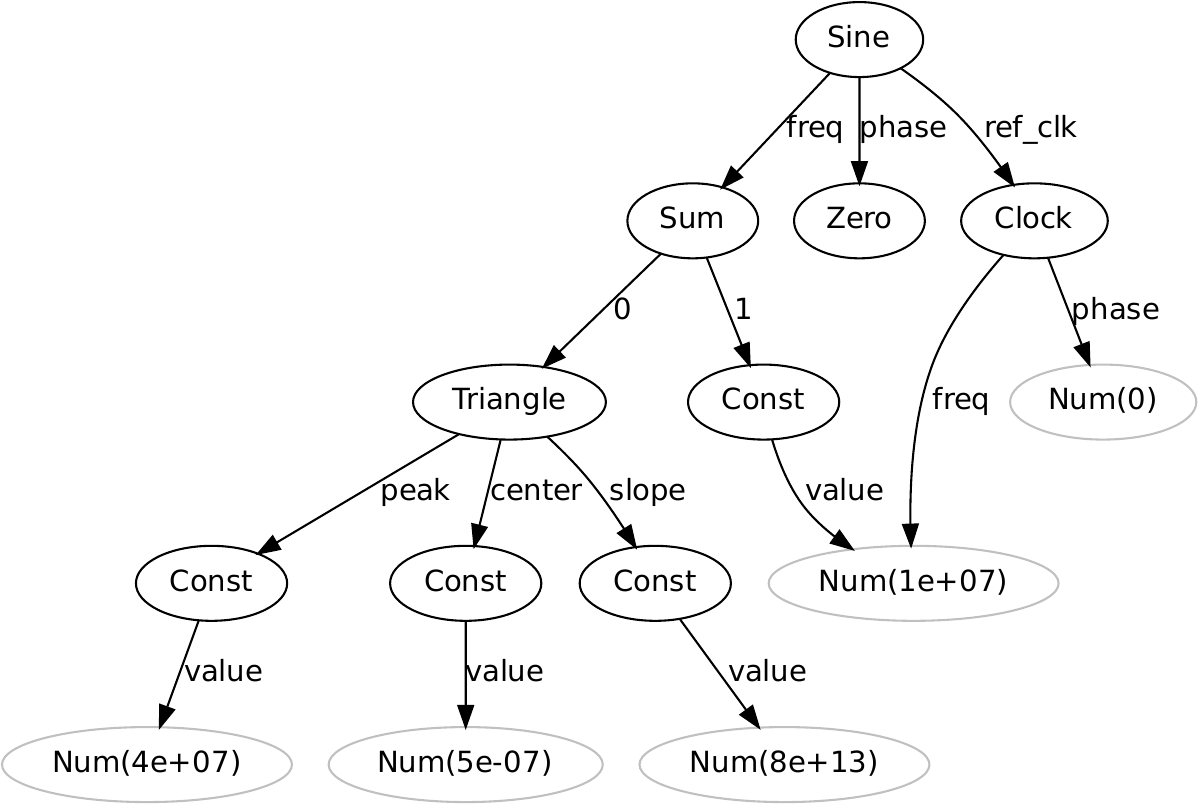}
    \caption{The same waveform as shown in Figure~\ref{fig:pulse:wave_sinefm}, but expressed using a regular sine node.}
    \label{fig:pulse:wave_sinefm_to_sine}
\end{figure}

Frequency and phase-modulated sine waveforms often use a carrier waveform and a modulation pattern that is relative to the carrier. We include specific waveform types for frequency and phase-modulated sine waveforms based on a carrier waveform, which are the sine FM and sine PM waveforms, respectively. Both waveforms have a carrier and modulation parameter where the carrier is a clock waveform and the modulation is relative to the carrier waveform.
For modulated sine waveforms, the carrier is always considered to be the reference clock.
Figure~\ref{fig:pulse:wave_sinefm} shows the graph of a sine FM waveform based on a 10~MHz carrier. The sine FM waveform can also be expanded to an equivalent regular sine waveform of which the graph is shown in Figure~\ref{fig:pulse:wave_sinefm_to_sine}. The sine FM graph and the regular sine graph clearly show how the sine FM node can compactly store carrier information, making it easier to interpret the graph. The sine FM and sine PM nodes are also examples of convenient semantic for creating modulated pulse sequences. Furthermore, the utility of the graph structure is highlighted by the ability to convert a sine FM node to an equivalent regular sine waveform which maybe required to realize the pulse for a specific target hardware or application.

\subsection{Schedule}
\label{sub:pulse:arch:schedule}

Waveforms allow us to describe individual pulses and sequentially timed pulses using the sequence waveform operator and zero waveforms. As a higher abstraction, \texttt{pulselib}'s API introduces \emph{channels} and a \emph{schedule} to provide convenient semantics for the description of multi-channel pulse schemes in the creation phase.

A channel is a unique identifier that represents an abstract analog output device. A channel has a label for representation purposes, but labels do not have to be unique. A user can create any number of channels, and it is up to the user to provide semantics for each created channel. In most cases, channels represent physical analog output devices, and a set of channels can be interpreted as an abstract device model. For example, a trapped-ion quantum computer might have a channel for the global beam and one individual beam channel for each ion, as shown in~\cite{kim2020hardware}.

A schedule is a map from channels to waveforms. All waveforms in a schedule have the same duration and are executed in parallel on their respective channels.
The schedule object provides utilities to construct valid multi-channel waveforms by providing a \emph{sequential} and \emph{parallel} context that can be nested in any arbitrary way.
We chose sequential and parallel semantics because a schedule can contain waveforms with variable duration. When the schedule is constructed, the parameters of waveforms, including the duration, might still be variable. Hence, it is not possible to evaluate the exact start and end times of waveforms. Using sequential and parallel semantics in combination with scalar operators allows users to construct practical schedules in the presence of variable waveform durations. Once all variables in a schedule are resolved, all waveform start and end times are known, and the schedule preserves its sequential and parallel semantics.

% Users can add waveforms to specific channels using the schedule object. By default, the schedule starts in a sequential context which is stored on the schedules' timing context stack. Users can enter and exit new contexts, which will be added and removed from the stack. Depending on the current context (i.e. the context on top of the stack), a waveform will be added to the schedule with sequential or parallel semantics. In a sequential context, the provided waveform is appended to the current waveform of the given channel, and a zero waveform of the same scalar duration is appended to all other channels in the schedule. In a parallel context, only one waveform can be added to each channel in the schedule. When another waveform is added to the same channel, the latest waveform will overwrite the existing one. The duration of a parallel context is defined as the maximum duration of its waveforms. When the parallel context is closed, a scalar max operator is used to obtain the duration of the context, and each waveform in the context is padded with a zero waveform to ensure all waveform durations are equal.
% When the user exits a timing context, it is removed from the stack and its contents are added to the new context on top of the stack.

By default, the duration of the sequential and parallel context scales dynamically based on the added waveforms. Dynamic scaling of context duration is convenient for situations where the duration between waveforms is expressed as the time between the end of one waveform and the start of another.
The sequential and parallel context can optionally be configured with a target duration given as a scalar. The target duration allows users to express the start time of one waveform relative to the start time of another.
% , which can be convenient for specific situations.
When a context is configured with a target duration, the context will add additional padding to each channel when the context is closed to meet the target context duration.
Configuring a target duration for a context can lead to invalid schedules, for example, if a waveform added to the context has a longer duration than the target context duration. Scheduling violations might not be known when the schedule is created due to variable waveform durations. Once all variables are substituted, scheduling violations can be recognized by waveforms with a negative duration.

\section{Transformations}
\label{sec:pulse:transform}

The \ac{dag} architecture for waveforms and schedules described in Sec~\ref{sec:pulse:arch} allows us to validate and transform the pulse representation using recursive graph algorithms. This empowers pulse optimization, transformation for target applications/hardware, and lowering the pulse to the realization phase.  In this section, we will discuss the graph visitor and transformer architecture and how they can be used to obtain the desired pulse format.

\subsection{Visitors and Transformers}
\label{sub:pulse:visitors}

Our graph algorithms are based on a visitor technique that recursively walks over the graph and calls a visit function for every node. Such a visitor technique is common and is, for example, used in the Python \ac{ast} module~\cite{python_ast}.
A visitor is a class with a \verb|visit()| method that takes a node as an argument and returns a node.
For each node that is visited, the visitor tries to find an appropriate visit method based on the class (i.e. type) of the node. When found, the class-specific visit method is called with the node as the argument. If no visit method is found, a generic visit method is used, which will call the \verb|visit()| method on all sub-nodes of the current node.
Parameters of a node and items of an operator node are considered sub-nodes in the \ac{dag} representation.
Finally, the \verb|visit()| method returns the original node.

We extend the visitor infrastructure by introducing type matching based on the class or superclasses of the node. Our matching algorithm first searches for a visit method for the class of the current node. If no match is found, the \verb|visit()| method will continue the search for a class-specific visit method using the superclasses of the node. Only if none of the node superclasses returns a match, the generic visit method is called.
For languages that support multiple inheritance (e.g. Python), superclasses are searched in the order determined by the C3 superclass linearization algorithm~\cite{barrett1996monotonic} (i.e. the \ac{mro} in Python).
Finally, we allow class-specific visit methods to reject a node which will cause the \verb|visit()| method to continue the search for a visit method.
Our visitor infrastructure implements a basic form of structural pattern matching where we only match the class and superclasses of the current node. More complex structural pattern matching, such as required for maximal munch~\cite{cattell1980automatic}, can be achieved by overriding the default behavior of the visitor.

To modify and transform a \ac{dag}, we use a transformer class which is an extension of the visitor class. A transformer works based on the same principles as the visitor, except that visit methods can return any node. The returned node can be the original node passed to the visit method to indicate that the node remains unchanged. If a different node is returned, the new node will replace the original node. The generic visit method of the transformer calls the \verb|visit()| method on all sub-nodes of the current node, and if any changes to the sub-node are detected, the current node is reconstructed with the new sub-nodes.

Visitors and transformers can be used to verify and transform user-provided pulse descriptions into representations that suit the needs of the target application. Transformations of interest include graph simplification and graph formatting to ensure the graph has a predetermined shape. Verification algorithms include graph validity checks and graph content checks. Such algorithms can be used to ensure the target application supports a given pulse description. Visitors and transformations always map graphs to graphs and are normally stateless (i.e. functional). Finally, visitors with linearization algorithms, such as maximal munch~\cite{barrett1996monotonic}, can convert graph-based pulse descriptions to linear data formats for realization by the target application and hardware. Visitors that convert graph-based pulse descriptions to other formats can use local data structures to generate output as a side-effect while traversing the graph. Hence, these visitors are not stateless and cannot be reused for a second pass without clearing their internal state. \texttt{Pulselib}, thereby not only helps with the creation and representation of the pulse but the transformers and munchers also allow the pulse to be appropriately converted and lowered to a format that can be realized for a target application on a specific hardware platform.

\subsection{Pipeline}

% A common use case for visitors and transformers is to take a user pulse description or schedule as input and use a sequence of visitors to transform the graph-based pulse descriptions to the desired target format.

\begin{figure}
    \centering
    \includegraphics[width=\linewidth]{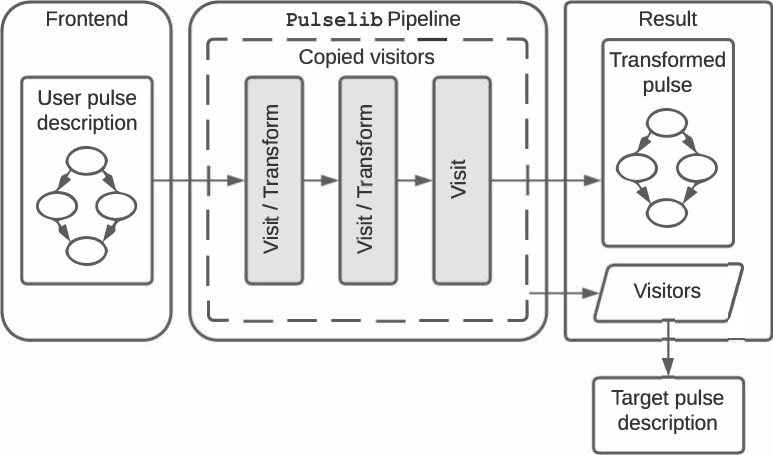}
    \caption{Schematic overview of a frontend and pipeline used to generate a target pulse description.}
    \label{fig:pulse:pipeline}
\end{figure}

We introduce a \emph{pipeline} to collect a sequence of visitors and apply them all sequentially on a pulse or schedule.
A pipeline contains a sequence of visitors (and transformers) and has a \verb|run()| method that accepts a single graph or a schedule. If we call \verb|run()| with a graph, the pipeline makes a copy of all visitors and applies them to the graph sequentially, where the output of the previous visitor becomes the input of the next. After the last visitor is applied, the pipeline constructs a result object, which is a tuple of the final graph and the visitors used to obtain the result.
When we call \verb|run()| with a schedule, the pipeline returns a map with channels and result objects as keys and values, respectively. The sub-pipeline for each channel is independent, and we can run their operations in parallel using common multiprocessing techniques.
The pipeline works very similar to the passes of a compiler where the graph functions as an \ac{ir}.
A schematic overview of a pipeline acting on a single graph is shown in Figure~\ref{fig:pulse:pipeline}.

% We would like to elaborate on the deep-copy behavior of the pipeline. Most visitors are stateless, which means the same object can be reused for multiple graphs or channels, but there are exceptions. As mentioned in Section~\ref{sub:pulse:visitors}, visitors with linearization algorithms can be stateful, and such visitors can not be reused without clearing their state. Instead of clearing the state of stateful visitors, we decided to create a copy of each visitor before it is applied to a graph. The state of visitors after traversing the graph is now part of the result. Hence, the visitors are part of the result object constructed for each graph that is processed by the pipeline.

\section{Implementation}
\label{sec:pulse:implementation}

We have implemented the pulse representation presented in Section~\ref{sec:pulse:arch} as a Python library, called \texttt{pulselib}~\cite{pulselib_git}, supporting Python version~3.8 and newer.
% . Our library is called \texttt{pulselib} and will be available as a publicly accessible project~\cite{riesebos2022pulseIR}.
The goal of \texttt{pulselib} is to serve as a pulse representation library that can be utilized as an interface or \ac{ir} for other software and research tools.
In this section, we will outline some of the design and implementation details of \texttt{pulselib}. These include details of semantics that allow for convenient pulse creation and the architecture that represents the pulse in memory. Additionally, we discuss various transforms and how we envision the usage of \texttt{pulselib}.

% @software{riesebos2022pulseIR,
%     title        = {pulseIR},
%     author       = {Riesebos, Leon and Brown, Kenneth R},
%     url          = {https://gitlab.com/duke-artiq/pulseIR},
%     year         = {2022},
% }

\subsection{Scalars and Waveforms}

From a graph perspective, there are only two fundamental node types: regular nodes and operator nodes. Regular nodes have labeled parameters nodes, and their relations are represented by labeled directed edges. Operator nodes have an ordered set of items where the relations are represented in the graph by enumerated directed edges.
At the same time, a node can either be a scalar or a waveform.
With Python as our host language, we chose a class structure based on multiple inheritance. For multiple inheritance, Python uses the C3 superclass linearization algorithm~\cite{barrett1996monotonic} to derive the \ac{mro} of a class, and we will use mixins to ensure our inheritance structure leads to a valid \ac{mro}.

% Our class structure starts at the root with an abstract \verb|BaseNode| class, which is the superclass of the (regular) \verb|Node| and the \verb|OperatorNode| class. The \verb|BaseNode| class defines an abstract method to iterate over sub-nodes, which can be used to easily walk over the graph independent of the node class. A \verb|Node| stores its labeled parameters in a dictionary where the keys are the strings of the labels and the values are node objects. An \verb|OperatorNode| stores its items in a list of node objects.
% To distinguish between scalar and waveform nodes, we introduce two abstract mixin classes \verb|Scalar| and \verb|Waveform|, which both inherit from the \verb|BaseNode| class. The mixin classes define generic abstract methods for their respective type, which is a \verb|value()| method for the \verb|Scalar| class and a \verb|duration()| method for the \verb|Waveform| class.
% % The \verb|Scalar| class defines a \verb|value()| method and the \verb|Waveform| class defines a \verb|duration()| method.

A concrete scalar node class is created by inheriting the \verb|Scalar| and the \verb|Node| class. For example, the number scalar class is defined as \verb|Num(Scalar, Node)|. The inheritance order is relevant for resolving the \ac{mro}, and the mixin should be the leftmost class in the inheritance order to get the desired \ac{mro}.
% The number scalar takes a float as an argument which is the value the node represents.
The variable scalar, defined as \verb|Variable(Scalar, Node)|, takes a key as an argument and has methods for substitution and clearing. The \verb|substitute()| method takes a dictionary that maps keys to values. The variable will look up its key in the dictionary and store the corresponding value. The \verb|clear()| method clears the memory that stores the value of the variable.
Concrete waveform node classes use an inheritance structure similar to scalars. 
% Hence, the constant waveform class is, for example, defined as \verb|Const(Waveform, Node)|.
% All waveforms listed in Table~\ref{tab:pulse:waveforms} are implemented using the same inheritance structure.

For operators, we introduce two intermediate abstract classes: \verb|ScalarOperator| and \verb|WaveformOperator|. Both classes inherit from their respective mixin class and the \verb|OperatorNode| class. The intermediate operator classes do not add any functionality to their subclasses but allow visitors to create specific visit methods for scalar and waveform operators based on the type-matching algorithm. 

% Scalar and waveform operators can be created by inheriting their respective operator class, for example, \verb|ScalarSum(ScalarOperator)| and \verb|Sum(WaveformOperator)|. The partial class diagram for nodes, scalars, and waveforms is shown in Figure~\ref{fig:pulse:class} which includes the concrete classes for \verb|Num|, \verb|Const|, \verb|ScalarSum|, and \verb|Sum|. The \ac{mro} for the same four concrete classes is shown in Table~\ref{tab:pulse:mro}.

% \begin{figure}
%     \centering
%     \includegraphics[width=\linewidth]{fig/pulseIR_class_diagram.png}
%     \caption{A partial class diagram of nodes, scalars, and waveforms.}
%     \label{fig:pulse:class}
% \end{figure}

% \begin{table}
% \centering
% \caption{The \acl{mro} of the four concrete scalar and waveform classes shown in Figure~\ref{fig:pulse:class}.}
% \begin{tabular}{@{}ll@{}}
% \toprule
% Class            & \Acl{mro} (superclasses only)\\
% \midrule
% \verb|Num|       & \verb|Scalar|, \verb|Node|, \verb|BaseNode| \\
% \verb|Const|     & \verb|Waveform|, \verb|Node|, \verb|BaseNode| \\
% \verb|ScalarSum| & \verb|ScalarOperator|, \verb|Scalar|, \verb|OperatorNode|,\\
%                  & \verb|BaseNode| \\
% \verb|Sum|       & \verb|WaveformOperator|, \verb|Waveform|, \verb|OperatorNode|,\\
%                  & \verb|BaseNode| \\
% \bottomrule
% \end{tabular}
% \label{tab:pulse:mro}
% \end{table}

The \verb|BaseNode| class and all its subclasses are immutable, similar to graph objects in functional languages. Immutable nodes allow graph algorithms to recognize modifications in sub-nodes or subgraphs by comparing object identities and guarantee that sub-nodes can not be mutated after a node is constructed. Any change in a leaf node will cause all of its super-nodes to be reconstructed, while subgraphs can be reused by multiple super-nodes.
% Additionally, we use Python \verb|__slots__| to prevent users from dynamically adding attributes to any node in the graph.
The \verb|Scalar| and \verb|Waveform| classes also overload a set of operators to allow users to create operator nodes using their respective operator syntax (i.e. \verb|a + b| for a scalar or waveform sum). Since nodes are immutable, in-place operators return new operator nodes too.
Implicit type casting allows objects of type \verb|int| or \verb|float| to be promoted to \verb|Num| scalars while all \verb|Scalar| types can in turn be promoted to \verb|Const| waveforms. When a type is implicitly promoted to a \verb|Const| waveform using an operator, the duration of the new waveform is set to the duration of the left operand. These are all examples of the semantics that allow for convenient pulse creation by \texttt{pulselib}.
% Figure~\ref{fig:pulse:operator_type_promotion} shows the resulting graphs of using an operator on a scalar or waveform with implicit type promotion on the right operand.

% \begin{figure}
%     \centering
%     \begin{subfigure}{0.9\linewidth}
%         \centering
%         \includegraphics[scale=0.45]{fig/scalar_operator.gv-crop.pdf}
%         \caption{\texttt{Num(value=5) + 3}}
%     \end{subfigure}
%     \\
%     \begin{subfigure}{0.9\linewidth}
%         \centering
%         \includegraphics[scale=0.45]{fig/waveform_operator.gv-crop.pdf}
%         \caption{\texttt{Const(value=5, duration=1e-3) * 2}}
%     \end{subfigure}
%     \caption{The resulting graphs of using an operator on a scalar or waveform with implicit type promotion on the right operand.}
%     \label{fig:pulse:operator_type_promotion}
% \end{figure}

\subsection{Schedule}

We develop a \verb|Channel| and a \verb|Schedule| class to create pulse schedules. The \verb|Channel| class only functions as an identifier to distinguish channels and additionally stores a non-unique label to provide a name for the channel. Besides the label, the channel object has no additional functionality.
The \verb|Schedule| class is a context and has an \verb|add()| method that takes a channel and a waveform as arguments.
When entering the schedule context, a time-context stack with a single sequential time context is created. The time context on top of the stack is considered the current context.
Users can request new sequential and parallel time contexts from the schedule object. The duration of a time context scales dynamically by default, but users can pass a fixed context duration if desired.
A time context is pushed to the stack when entered and removed when exited.
The \verb|add()| method of the schedule object forwards the call to the current time context, which will decide its timing interpretation and store the information in a dictionary with channels and waveforms as keys and values, respectively.
When a time context is removed from the stack, its contents are extracted and added to the time context now at the top of the stack.
Interpretation and details of the sequential and parallel time context are outlined in Section~\ref{sub:pulse:arch:schedule}.
Note that a schedule never tries to resolve the duration of a waveform at any time and instead uses scalar operators to determine the durations of padding waveforms. Once all durations are known, a transformation can easily remove any redundant padding nodes.
When the user exits the schedule context, the initial sequential context is closed, and its waveform dictionary, which represents the full schedule, is extracted. The schedule stores the dictionary, and users can request the waveform dictionary from the schedule object.
% Listing~\ref{lst:pulse:schedule} shows a code snippet for creating a simple two-channel schedule, and Figure~\ref{fig:pulse:schedule_plot} shows the resulting waveforms.

% \begin{listing}
%     \inputminted{python3}{lst/schedule.py}
%     \caption{Snippet of a two-channel schedule with sequential and parallel waveforms.}
%     \label{lst:pulse:schedule}
% \end{listing}

% \begin{figure}
%     \centering
%     \includegraphics[width=\linewidth]{fig/wave_schedule.pdf}
%     \caption{The rendered waveform of the schedule shown in Listing~\ref{lst:pulse:schedule} using a sample rate of 1~GHz.}
%     \label{fig:pulse:schedule_plot}
% \end{figure}

% \subsection{Graph and Pulse Rendering}

% For visualization purposes, we have implemented a recursive algorithm that creates a GraphViz file from a pulse description. We use the Python GraphViz library to create a \verb|*.gv| file which can be rendered using the GraphViz \verb|dot| program for \acp{dag}. All graph figures in this chapter have been created with GraphViz.

% All waveform classes implement a \verb|to_array()| method that recursively calculates the samples of a waveform given a sample rate. The \verb|to_array()| method can be used to target an \ac{awg} device, export waveforms to a standard \ac{vcd} file using PyVCD, or plot waveforms using a tool like Matplotlib. All pulse plots in this chapter have been created with Matplotlib.

\subsection{Visitors and Transformers}

% We created base classes for visitors and transformers that implementations of graph algorithms can inherit.
The abstract \verb|BaseVisitor| class contains an abstract \verb|visit()| method that takes a \verb|BaseNode| as an argument and returns a \verb|BaseNode|. The concrete \verb|Visitor| and \verb|Transformer| classes inherit from the \verb|BaseVisitor| class.
As outlined in Section~\ref{sec:pulse:arch}, the \verb|visit()| method of the \verb|Visitor| class implements a basic type-matching algorithm that obtains the \ac{mro} from the node's class and iterates over the classes to find a matching visit method. The class-specific visit method is found by dynamically searching for a method \verb|visit_*()| where the asterisk represents the (case-sensitive) name of the target class. If a class-specific visitor method is found, the method is called with the current node as an argument. 
% The class-specific visit method can raise a \verb|ContinueVisit| exception, in which case the search for a class-specific visit method continues.
If no appropriate visit method was found, the current node is passed to the \verb|generic_visit()| method. The generic visit method continues the graph traversal by calling the \verb|visit()| method on all sub-nodes. By default, the \verb|Visitor| class visits all nodes depth-first.
Finally, the \verb|visit()| method returns the original node that was passed as an argument.
% The implementation of the \verb|Visitor| class, including the \ac{mro} traversal and the \verb|ContinueVisit| handling, is shown in Listing~\ref{lst:pulse:visitor}.
% More complex structural pattern matching can be achieved by implementing a \verb|visit()| method that uses the Python \verb|match| statement (Python 3.10 or newer as specified in PEP~634~\cite{pep634}) or any third-party library that achieves similar functionality.

% \begin{listing}
%     \inputminted[firstline=26,lastline=54]{python3}{lst/visitor.py}
%     \caption{The visitor implementation.}
%     \label{lst:pulse:visitor}
% \end{listing}

The \verb|Transformer| class uses the same type-matching algorithm as the \verb|Visitor| class, but additionally allows modifications to nodes in the graph.
If the returned node object is the same as the original node provided to the visit method, the node remains unchanged. If a different node object is returned, it will replace the original node.
As a result of node immutability, visited nodes in the graph must be reconstructed if sub-nodes are modified. Hence, there is no generic visit method that accepts every type of node. Instead, the transformer class has two generic class-specific visit methods for the \verb|Node| and \verb|OperatorNode| classes. These two fallback methods cover all node types and can reconstruct node objects if any sub-nodes are modified.
% The implementation of the \verb|Transformer| class is shown in Listing~\ref{lst:pulse:transformer}.
To ensure nodes can be automatically reconstructed if sub-nodes are modified, class-specific visit methods for scalars and waveforms must return scalar and waveform nodes, respectively. A visit method for a waveform could theoretically return a scalar due to implicit type promotion, but the unconstrained duration could cause unintended side effects. In most cases, a clock waveform can neither be replaced with another waveform type. If any transformation results in an illegal waveform or scalar, the transformer will fail and raise an exception.

% \begin{listing}
%     \inputminted[firstline=58,lastline=104]{python3}{lst/visitor.py}
%     \caption{The transformer implementation.}
%     \label{lst:pulse:transformer}
% \end{listing}

% At the moment of writing, we have already implemented a limited set of visitors and transformers, and more are in development. We will discuss a few of them in this paragraph.
Two implemented visitors are for substitution and clearing of variable scalars. The substitution visitor takes a single dictionary and uses it to call the substitute method of every variable in the graph. The clearing visitor calls the clear method of each variable instead.
% As mentioned in~Section~\ref{sub:pulse:arch:schedule}, we also developed a visitor that verifies that no waveform duration is negative to ensure waveforms or schedules are valid.
Other common transformations cover the simplification of scalar and waveform graphs. Such transformations include variable substitution, folding, and operator simplification. Simplification transformations are most effective when all variables are substituted, but can also be effective with unsubstituted variables.
% For example, the schedule object inserts a lot of padding nodes that can easily be simplified once waveform durations are substituted.
Finally, we develop maximal munch visitors that can be used to transform the pulse representation into a linear data format. Maximal munch visitors often only match a sequence of specific sub-graphs supported by the target output, and any unmatched item will cause an exception. A pattern mismatch indicates that the graph contains waveforms that are not supported by the target output. Hence, maximal munch visitors not only linearize the data but also verify waveforms are supported by the target.

\section{Applications}
\label{sec:pulse:applications}

We demonstrate the utility of \texttt{pulselib}'s \ac{dag} architecture and feature set through some applications involving phase synchronization of pulses. We describe two common examples in quantum computation that require accurate phase tracking.  The first involves a time-dependent carrier resonance that depends on the specific pulse we apply, and we overcome this with \texttt{pulselib}'s clock sequence.  The second is the use of more than two quantum states to create a qudit. This results in the need for more reference clocks because there is more than one resonance frequency.  Although we will utilize trapped ions as the example platform for these situations, the issues are general to all platforms.

\subsection{Pulse-Dependent Clock}
Although qubits are ideal two-level systems, in practice all platforms have more states available.  When applying a pulse to couple the qubit states together, it is inevitable that the qubit states are coupled to these auxiliary states as well. Assuming this coupling is far-detuned, leakage errors will be avoided but there will be state dependent energy shift on the system. This energy shift, often referred to as an AC-Stark shift, results in a shift in the resonance of the qubit and in the frequency of the reference clock \cite{schuster2005ac, haffner2003precision}. 

To overcome this issue, \texttt{pulselib} allows for the use of clock sequences.  Each pulse can reference a clock associated with its specific resonance, and the phase of that clock can be connected to a previous pulse's clock using this sequencing technique.

As an illustrative example, let's consider Raman transitions in hyperfine trapped ions \cite{wineland1998experimental}. Single qubit gates are performed using two counter-propagating lasers whose absolute frequency is detuned from a very strong dipole transition, where a set of high-energy auxiliary state are present.  The beatnote of these two lasers make up the frequency difference of the qubit states.  The presence of the auxiliary states acts as a virtual bridge for the qubit population to switch states.  However, the presence of these states causes the aforementioned AC-Stark shift that changes the clock frequency \cite{lee2016engineering}.  There is thus a different reference clock frequency when a pulse is being applied versus when the qubit is idling.   

To further complicate things, the most common two-qubit gate is the \ac{ms} gate, and it requires three lasers on each ion \cite{sorensen1999quantum, sorensen2000entanglement, molmer1999multiparticle}.  The presence of the third laser allows for a spin-dependent coupling to the ion's harmonic motion: one laser creates a beatnote with a second that is the qubit resonance plus the harmonic frequency, and the final laser provides a similar beatnote with the second that is the qubit resonance minus the harmonic frequency.  However, this creates a different AC-Stark shift than the single qubit case.  

In the simplest case, there are at most two total \ac{rf}/\ac{mw} sources present in all three cases (although in practice there can be many more): zero in the idling case, one in the single qubit case to create the beatnote, and two in the two-qubit case to create the two different beatnotes. To simplify the two-qubit case, we will assume one \ac{rf}/\ac{mw} tone creates the qubit resonance beatnote, and the second tone mixes with this one to create the harmonic motion beatnote. This simplifies the sequencing greatly because it allows one \ac{rf} tone to be associated with the carrier transition, which is the phase we care about.  

\begin{figure*}
    \centering
    \includegraphics[width=\linewidth]{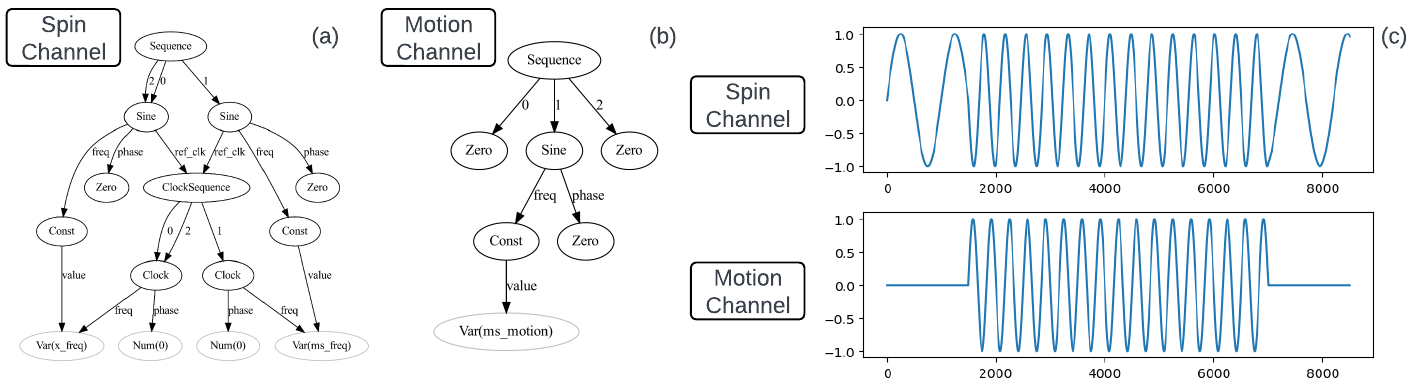}
    \caption{Pulse sequence for a pulse dependent clock, in particular a single qubit gate followed by a two qubit \ac{ms} gate followed by another single qubit gate. (a) The representation of the spin channel pulses.  This channel tracks the phase of the qubit itself and is used by both single qubit and two qubit gates. This is where the change in frequency due to different pulses is taken into account with a clock sequences. (b) The representation of the motion channel, which tracks the motion of the trapped ion. The phase of this channel doesn't matter for standard quantum computation with trapped ions, and thus no clock is referenced. (c) Visual representation of the change in frequency in the clock due to the different pulses using toy numbers.  As is shown, the spin channel changes frequency during single and two qubit gates.  The motion is only used during the two qubit gate.}
    \label{fig:pulse:ms_gate}
\end{figure*}

A pulse schedule that properly tracks all of the relevant phases is shown in Figure \ref{fig:pulse:ms_gate}.  The \ac{rf}/\ac{mw} tone that creates the qubit resonance is labeled spin channel, and the one responsible for motional control is labeled motion channel.  The use of the clock sequence is best demonstrated in Figure~\ref{fig:pulse:ms_gate} (c), where the change in frequency due the two qubit pulse is demonstrated using toy numbers.

\subsection{Shelving and Qudits}
The presence of auxiliary states also means we can use them for something practical, whether as a means to shelve during an intermediate measurement, or as more states to increase computational power \cite{allcock2021omg, edmunds2021scalable, goss2022high, neeley2009emulation, ringbauer2022universal}.  In hyperfine trapped ions for example, this could be the magnetic field sensitive Zeeman states or metastable states in the $D$ manifold.  Regardless of the choice of states, there will be a different resonance frequency with each transition we wish to perform.  This means there is a different reference clock with each transition as well.  

\begin{figure*}
    \centering
    \includegraphics[width=\linewidth]{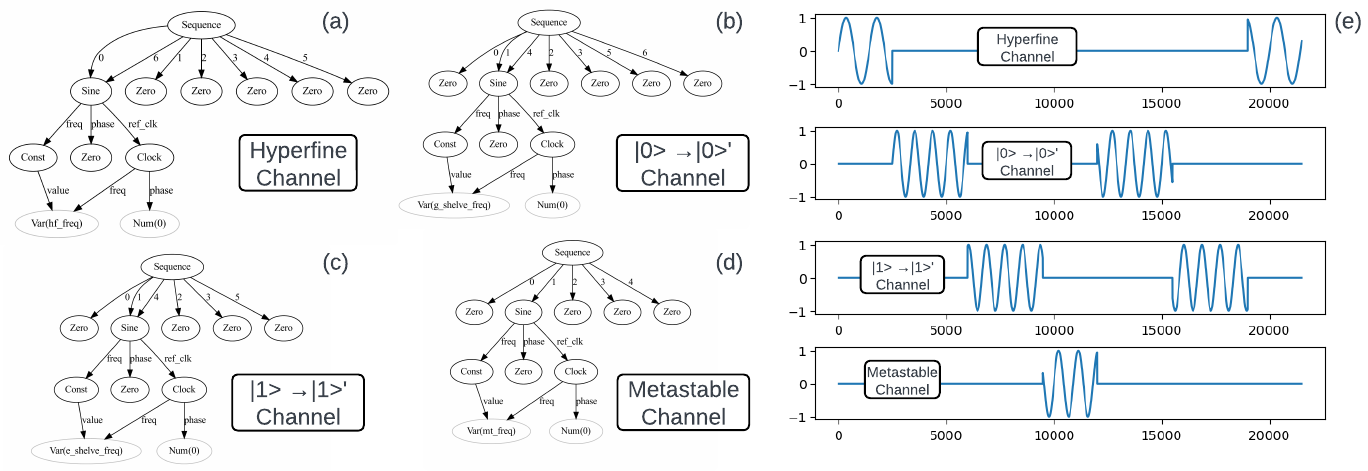}
    \caption{Pulse sequence for shelving in a trapped ions system using hyperfine ground states (labeled $\ket{0}$ and $\ket{1}$) and metastable auxiliary states (labeled $\ket{0'}$ and $\ket{1'}$). The sequence consists of a $\sqrt{\sigma_y}$ in the hyperfine states, followed by transfer from hyperfine to metastable states, followed by a $\sigma_y$ in the metastable states, followed transfer back to the hyperfine states, followed by a final $\sqrt{\sigma_y}$ pulse.  The final state should be the same as the input state if every pulse is coherent. (a)-(d) Pulse representations of the four required channels during this pulse sequence.  Note that each of them reference a different clock. (e) Visual representation of the clocks of each pulse.  Note that none of them change frequency, and as long as they reference the same absolute phase as each other at the start time, then there is no tricky phase tracking.}
    \label{fig:pulse:shelving}
\end{figure*}

This problem can be overcome quite easily using different reference clocks in \texttt{pulselib}.  Let's take the hyperfine ion as an example, where we choose a four state system defined by two states in the $S$ manifold, labeled $\ket{0}$ and $\ket{1}$, and two more in the metastable $D$ manifold, labeled $\ket{0'}$ and $\ket{1'}$. We will start in the state $\ket{0}$ and perform a $\sqrt{\sigma_y}$ pulse to move to $\ket{+}$ = $\frac{1}{\sqrt{2}}(\ket{0} + \ket{1})$. We choose this state because it is the most sensitive to phase noise. Pulses between these two states define one reference clock, which will be in the \ac{rf}/\ac{mw} regime and thus the phase is easily controllable. We then perform a pulse to move $\ket{0}$ to $\ket{0'}$ and $\ket{1}$ to $\ket{1'}$.  These two pulses have their own separate reference clocks. However, we assume the same laser performs each pulse, and the frequency difference is made up for with two different \ac{rf}/\ac{mw} sources. This assumption is important because it allows us to ignore a global laser phase that is imparted onto the qubit and instead focus on the \ac{rf}/\ac{mw} phase. Next, we perform another $\sigma_y$ in the $D$ manifold to move into the $\ket{-'}$ = $\frac{1}{\sqrt{2}}(\ket{0'} - \ket{1'})$ state.  This pulse has its own reference clock as well, also in the \ac{rf}/\ac{mw} regime. Finally, we move back to the ground state and perform another $\sqrt{\sigma_y}$ to move back to the beginning state $\ket{0}$. It can be shown that as long as we can set all of the \ac{rf}/\ac{mw} sources to the same absolute value at the beginning of the experiment, then we can allow each of these clocks to run unchanged throughout the experiment (i.e. no need for clock sequences).  We just need to reference each of these clocks using \texttt{pulselib} when applying the appropriate pulses, as demonstrated in Figure \ref{fig:pulse:shelving}.

Finally, it is worth noting that state-dependent clock shifts will happen in this situation as well.  To account for this, we would need to take advantage of the clock sequences that \texttt{pulselib} provides in conjunction with having multiple clocks.
\acresetall
\section{Conclusion}
\label{sec:pulse:conclusion}

% We have presented a pulse description architecture that efficiently stores high-level pulse descriptions in a graph-based format. Our pulse representation supports arbitrary waveforms, but works most effectively with parameterized basic waveforms. Multi-channel pulse descriptions are supported by the channel and schedule infrastructure.
% Pulse descriptions can be transformed by applying recursive algorithms to the graph representation. Graph algorithms can be implemented using our visitor infrastructure using class and subclass-based type matching and pattern matching techniques. With the help of pipelines, pulse schedules can be transformed easily into linear data formats suitable for simulation or execution on a target device.
% All our proposed techniques are implemented in Python as part of our pulselib library.
% Finally, we have outlined our vision for integrating pulselib into quantum simulators and real-time quantum control systems.

With a large number of high-level pulse representations, we present a unique pulse architecture that efficiently stores high-level pulse descriptions in a graph-based format. It provides semantics for convenient and complete pulse \emph{creation}, provides efficient parameter-based \emph{representation} that retains information from the \emph{creation} phase, and allows for \emph{realization} to a target application and hardware through transformations and visitors.  

Our pulse representation consists of parameterized basic waveforms, allows for creating arbitrary waveforms, and supports operations on these waveforms. The channel and schedule infrastructure supports multi-channel pulse descriptions.
Pulse descriptions can be transformed by applying recursive algorithms to the graph representation. Graph algorithms can be implemented using our visitor infrastructure using class and subclass-based type-matching and pattern-matching techniques. With the help of pipelines, pulse schedules can be transformed easily into linear data formats suitable for simulation or execution on a target device.
We demonstrate \texttt{pulselib}'s utility using motivating applications of pulse schemes used in trapped-ion quantum computers~---~AC-stark shift in gate operations and qubit shelving. In both these cases phase synchronization across operation is vital to accurately change the qubit's state. \texttt{Pulselib}'s graph architecture allows for the description and representation of phase synchronized pulses by allowing waveforms to track their phase using a reference clock, represented using clock waveforms.

% We have implemented \texttt{pulselib} as a Python library.

% \section*{Acknowledgment}
\section*{Acknowledgment}
The work was funded by the National Science Foundation (NSF) STAQ Project (PHY-1818914, PHY-2325080), EPiQC - an NSF Expeditions in Computing (CCF-1832377), NSF Quantum Leap Challenge  Institute for Robust Quantum Simulation (OMA-2120757), and the U.S. Department of Energy, Office of Advanced Scientific Computing Research QSCOUT program. Support is also acknowledged from the U.S. Department of Energy, Office of Science, National Quantum Information Science Research Centers, and Quantum Systems Accelerator. We thank Jonathan M. Baker for his input on this work.
\clearpage
\bibliographystyle{IEEEtran}
\bibliography{main}

\end{document}